# Complete inclusion of bioactive molecules and particles in polydimethylsiloxane: a straightforward process under mild conditions

Greta Faccio[a*], Alice Cont[b], Erik Mailand[a], Elaheh Zare-Eelanjegh[a], Riccardo Innocenti Malini[b], Katharina Maniura-Weber[a], René Michel Rossi[b] and Fabrizio Spano[b*]

[a] Empa, Swiss Federal Laboratories for Materials Science and Technology, Laboratory for Biointerfaces, Lerchenfeldstrasse 5, CH-9014 St. Gallen, Switzerland.

[b] Empa, Swiss Federal Laboratories for Materials Science and Technology, Laboratory for Biomimetic Membranes and Textiles, Lerchenfeldstrasse 5, CH-9014 St. Gallen, Switzerland.

*E-mail: fabrizio.spano@zhaw.ch

greta.faccio@gmail.com

**Abstract**

By applying a slow curing process, we show that biomolecules can be incorporated via a simple process as liquid stable phases inside a polydimethylsiloxane (PDMS) matrix. The process is carried out under mild conditions with regards to temperature, pH and relative humidity, and is thus suitable for application to biological entities. Fluorescence and enzymatic activity measurements, show that the biochemical properties of the proteins and enzyme tested are preserved, without loss due to adsorption at the liquid-polymer interface. Protected from external stimuli by the PDMS matrix, these soft liquid composite materials are new tools of interest for robotics, microfluidics, diagnostics and chemical microreactors.

Liquids and elastomeric materials are key components in microfluidic and diagnostic devices. Conveyed in open structures such as microchannels[1], deposited as refillable reservoirs[2], or conductive metallic lines[3], liquids can add novel functionalities to materials. For instance, in closed structures such as droplets, liquids reinforce the mechanical properties of elastomeric composite material[4]. Our work reports a novel inclusion process and addresses how it affects the molecules contained in the droplets, in particular bioactive molecules or particles embedded in a polydimethylsiloxane (PDMS) matrix.

The introduction of compartments, channels and cages in materials is a challenging, yet desired step towards the manufacture of complex lab-on-a-chip architectures, metamaterials[5], flexible sensors[6], stretchable electronics[3], soft composites[7], mobile structures[8], and soft robots[9]. Compartmentalization of biomolecules and/or reactions benefit from the confinement in an isolated and thus contamination-free environment. When coupled to a high-throughput manufacturing strategy, this allows access to a large number of samples, each with different reaction conditions[10]. For instance, this has been used for screening and for crystallization studies, where nanoliter-scale droplets have been superficially deposited on a PDMS chip to imprint nanowells[11]. A similar approach has been applied to single protein counting[12] and single-cell analyses[13,14].

Droplets are complex systems characterized by a high internal mobility of the component molecules[15] and 'active' droplets able to self-divide, self-replicate, or migrate along chemical gradients can be assembled by tailoring their composition and surface tension[15]. Single size-controlled droplets provide microreactors for chemical reactions in a partially isolated environment, which is of high interest for microfluidic devices[16–18], or the analysis of reactions in confined spaces[19]. The inclusion of a liquid or gel phase in a polymeric material allows modification of the mechanical properties of the material and the generation of soft liquid composites (SLC) with novel properties and functionalities, e.g. for diagnostics[20–22], production of plasmonic devices[23], stretchable electronics[3], soft conductive three-dimensional sponges[24], and polymer-encapsulated liquid droplets[25].

Regardless of its high hydrophobicity (water contact angle of 110°[26]), PDMS is a versatile elastomeric polymer, easy to manipulate, with a very good contour accuracy (<10 nm), chemical inertness, thermal stability, and homogeneity even after the generation of microstructures[27,28]. Applications range from electronics to biomedical devices, mechanobiology and microfluidics. PDMS has been applied to the development of patches for the release of therapeutic molecules triggered by a mechanical stress[29], e.g. stretching, pressure. PDMS is often used in the prototyping of microfluidic devices but the small dimensions of the channels and the high hydrophobicity of the material makes it prone to biofouling, thus requiring passivation methods[30–36].

Current approaches to the production of microcompartments for liquids in elastomeric materials require laborious two-to-many step processes involving the deposition of multiple layers, the design of masks or moulds, lithography or replica molding[37]. In addition to increasing time and cost, multiple steps increase the risk of fabrication errors and possible contamination.

Here, we report a straightforward method for the inclusion of biomolecules as liquid droplets in PDMS films to generate SLC materials. In comparison to previously reported approaches, droplets are fully included in the PDMS matrix, and the formation and storage of biomolecules in aqueous droplets is achieved without surfactants or detergents that might compromise their bioactivity. Ensuring the stability of fluorophores and of catalytic entities is a requirement when testing how synthetic devices affect biological entities. Therefore, we firstly optimized the inclusion method by using dyes and then demonstrated its functionality by investigating three fluorescent proteins and the biotechnologically-relevant catalyst laccase. Additionally, we show that functional materials such as magnetic nanoparticles in liquid droplets can also be included in PDMS films and their spatial distribution can be controlled via external stimuli.

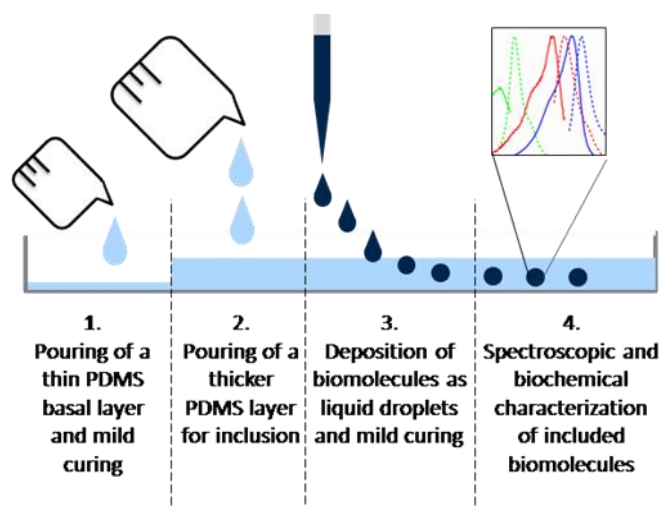

**Figure 1.** Schematic representation of the process leading to the inclusion of an aqueous solution containing a biomolecule in its active form in PDMS. After pouring a thin basal layer of liquid PDMS that is then cured, a subsequent thicker layer is added on top. Then the biomolecule-rich solution is deposited in µl-sized droplets and the PDMS is let to cure under mild condition, to be characterized subsequently by in situ spectrophotometry or ex situ using biochemical methods.

The initial hypothesis was that PDMS in its viscous state would be suitable for the inclusion of liquid droplets, which could introduce additional functionalities to the material. Two main issues had however to be faced. First, the high viscosity of unpolymerized PDMS prevents the spontaneous complete penetration of water-only liquid droplets, as their density is too low. Secondly, droplets with higher density dispensed at the surface of the unpolymerized PDMS slowly migrated towards the bottom and adhered to the mould used for casting, leading to a loss of integrity. To solve the first technical issue, the density of the liquid droplet was tuned to promote its complete incorporation in the PDMS matrix. Preliminary screening experiments identified glycerol as the optimal compound. Characterized by a higher density than PDMS ($d_{glycerol}$ = 1.26 g/cm³ and $d_{PDMS}$ = 0.97 g/cm3 at 20˚C), glycerol is able to stabilize the tertiary structure of biomolecules, it is used as cryoprotectant, inhibitor of aggregation, and is an inert component in enzymatic, biological, and chemical reactions[38]. Moreover, PDMS and glycerol were selected for their incompressibility and immiscibility. Therefore, glycerol was added at a concentration of 50% to all aqueous samples prior to deposition in PDMS. These droplets were then introduced within the PDMS matrix as shown in Figure 1. To ensure the integrity of the droplet within the material, first a basal layer of PDMS (0.5-1 mm thick) was poured into the mould and cured under mild conditions (Figure 1, step 1), e.g. room temperature (22°C) for 48 h. Once cured, a second unpolymerized PDMS layer was dispensed to reach an approximate 5:1 vol/vol ratio with the basal layer (Figure 1, step 2). As an alternative to degassing, air inclusions were let to spontaneously migrate to the surface and, when no longer visible (after approximately 30 min), droplets of the chosen biomolecule-containing solutions were dispensed on the surface of the viscous PDMS using a micropipette and let to settle (Figure 1, step 3). The deposition of the droplets was realized before the polymerization of the PDMS was complete, e.g. during an experimentally defined window of time of a few hours where the PDMS was still in a liquid viscous state, even if the polymerization process already started.

As an illustration of the final composite, Figure 2 shows the generation of aqueous dye-loaded inclusions into the PDMS matrix. The aqueous solution containing the blue-violet dye in distilled water was mixed with glycerol (50% volume of the droplet solution) to generate the droplets included into the PDMS matrix. Figure 2a and 2b illustrate the top view of the PDMS film with blue-violet droplets having a unique size. The droplets' volume was 10 µl and the droplets were deposited on a hexagonal lattice, demonstrating control over the localization of each droplet. Figures 2c and 2d indicate that the droplets were not on the surface of the PDMS matrix but were included inside the PDMS matrix as illustrated in Figure 2d where a side view of the PDMS matrix is shown.

In the next example (Figure 2e, 2f, 2g, and Figure SI1), droplets of different volumes were included in the PDMS to demonstrate control over their size. The volumes used were 5, 10, 15 and 20 µl. Figures 2e, 2f and 2g clearly indicate the possibility to control the volume and thus the size of the generated inclusions into the PDMS matrix allowing to tune both the material and the droplets/microreactors.

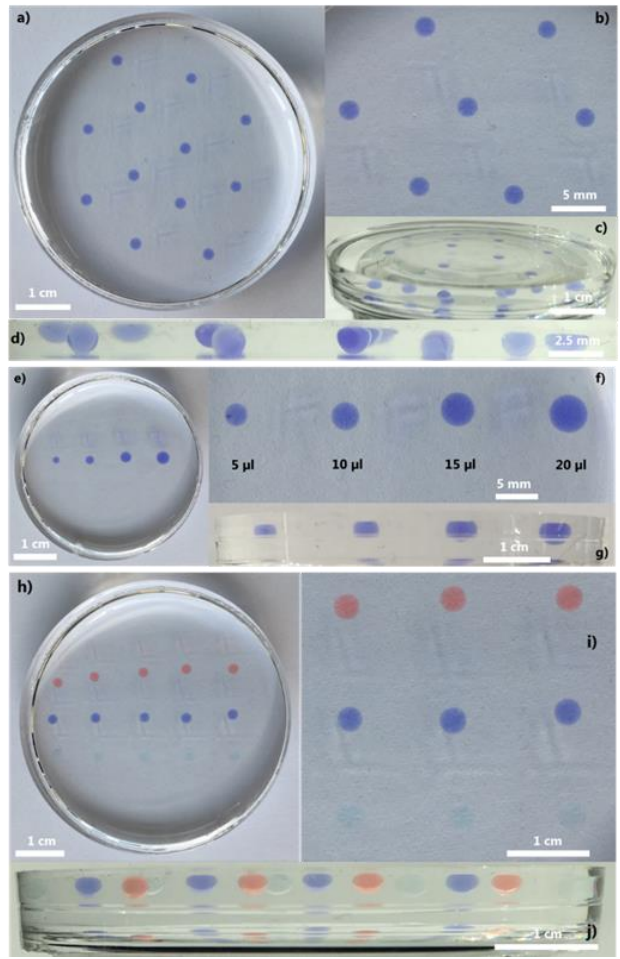

**Figure 2.** Soft liquid composite samples made with inclusions of various dye-loaded droplets: a) Top view of the blue-violet dye-loaded droplet inclusions forming a hexagonal network; b) Zoomed view of the hexagonal network; c) 60° and d) 90° side views of the sample illustrating the complete inclusion of the blue-violet liquid droplets into the PDMS matrix; e) Top view of inclusions made using droplets of 5, 10, 15 and 20 µl ; f) Zoomed view of the soft liquid composite sample illustrating the different droplets; g) 90° side view of the sample illustrating the penetration of the blue-violet liquid droplets into the PDMS matrix. h) Top view of the sample made with different dye-loaded droplets forming a cubic network; i) Zoomed view of the cubic network; j) 90° side view of the sample illustrating the penetration of the different dye-loaded droplets into the PDMS matrix.

As a last illustration of the developed method, in Figure 2h, 2i, and 2j, the introduction of aqueous droplets using different dyes was investigated. A cubic network was used to deposit the droplets. The droplets contained the same volume (10 µl) but were made of mixtures of different dyes in distilled water and 50 v/v % glycerol. Figure 2h and 2i display the top views of the sample. In particular, Figure 2i shows a zoomed view of the cubic network. In Figure 2j, we show the 90° side view illustrating the droplets into the PDMS matrix. In Figure SI2, additional illustrations indicate the change of colour in function of the view angles due to the superposition of the different dye-loaded droplets demonstrating that the material could be used to manipulate incoming light and potentially as a signal transducer.

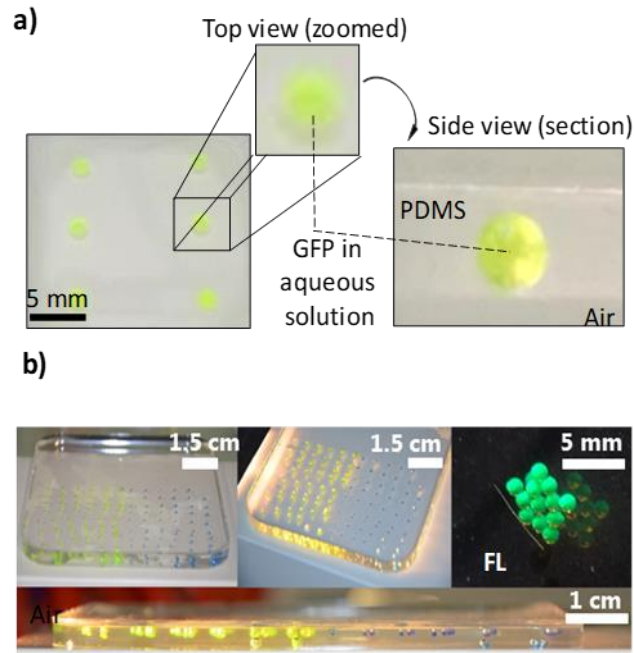

**Figure 3.** Inclusion of biomolecules containing droplets in a controlled spatial arrangement. (a) Zoomed view of inclusions of GFP in PDMS as droplets in liquid form. Deposited as 6 droplets of 10 µl volume, aqueous solutions containing GFP can be stored in PDMS as well-defined spherical droplets. (b) Droplets can be freely arranged in patterns and arrays without leading to fusion while preserving their fluorescent properties (FL).

After demonstrating control over the droplet distribution and size (Figure 2 and 3) fluorescent proteins were used to investigate the effect of the inclusion process on biomolecules (Figure 3a). Fluorescent proteins CPC, GFP, and mCherry were selected because of their different structural and fluorescent properties. Since the fluorescence of a protein is dependent on the interaction of the chromophore with its molecular environment, including the protein matrix[39], it can be perturbed by the interaction with a material surface[40,41]. It is thus interesting to investigate if the inclusion of fluorescent proteins solutions in a PDMS matrix affects their properties[42]. All fluorescent proteins used are characterized by good solubility levels and their molecular surface show well distributed hydrophobic regions that, however, could lead to interactions with the PDMS surface (Figure SI3). Although the adsorption of proteins to surfaces is a spontaneous process hard to prevent, and their interaction with PDMS surfaces in aqueous solutions was reported to lead to extensive adsorption forming layers of up to 4 nm in thickness with a density of ~3 mg/m2 in less than one hour, droplets containing the fluorescent proteins were homogeneous. The colour was not concentrated at the PDMS interface nor at the bottom of the droplets, which could happen in case of aggregation. Therefore, this suggests that the addition of the glycerol inhibits the adsorption of the proteins to the surface and minimizes their interactions with one another (Figure 3a). Upon inclusion and by controlling the direction of the white-light illumination, the characteristic fluorescence of the proteins became visible (Figure 4a, Figure SI4). Under lateral illumination, the included droplets of GFP and CPC appeared green and red to the eye, respectively, as these conditions reduced the amount of reflected white light observed under top illumination. CPC, GFP and mCherry proteins within droplets retained their characteristic light absorption and fluorescence properties (Figure 4b) once included, i.e. with absorbance/fluorescence maxima at 620/640 nm, 395/509 nm, and 587/610 nm, respectively. In addition to being measured in situ with a fluorimeter, fluorescence intensity was quantified using a microarray scanner (Figure 4c) and the linear correlation between concentration and emitted fluorescence of GFP was not affected by the inclusion process (Figure 4c) in a 0-2 mg/ml range. These results show that PDMS could be used as an optical waveguide[43] (Figure 4a), due to the intrinsic smoothness of the polymerized PDMS-liquid interface that minimizes light scattering and reflection. This might be of interest for static optofluidic applications, e.g. liquid lenses and mirrors.

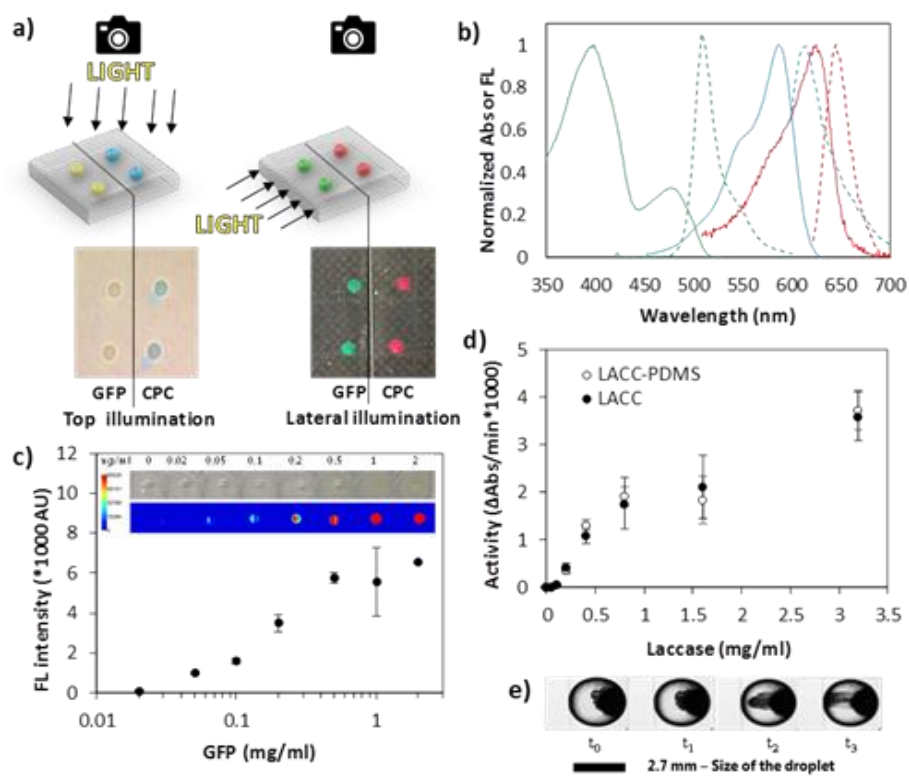

Figure 4. Characterization of fluorescent proteins, enzymes, and nanoparticles included as droplets in PDMS. (a) Fluorescence of the included biomolecules was visible upon lateral illumination. (b) In situ measurements of the absorbance (continuous line) and fluorescence (dashed line) emission spectra of GFP (green), mCherry (blue), CPC (red) and included as droplets in PDMS. (c) Concentration-dependent fluorescence emission of GFP as droplets in PDMS. In the inset, a photographic image and a false colouring based on fluorescence of the material. (d) Concentration-dependent activity of laccase included in PDMS (empty dots) and of the control solution (filled dots). (e) Migration of nanoparticle aggregates in aqueous solution as droplets when exposed to an external magnetic field (magnet on the left) and visualized by optical microscopy over time (droplet volume: 10 µl, = 2.7 mm, total time = 30s).

The enzyme laccase has proven to be particularly valuable in biocatalysis, food engineering, and sensing devices[44], it would therefore be interesting to be able to screen its properties in the presence of different solutes in a controlled environment. Like the fluorescent proteins used above, laccase is also a highly soluble protein that, additionally, is glycosylated at multiple sites, which contribute to its stabilization[45]. The interaction of proteins, applied in dry powder form, through hydrophobic and moulding interactions with unpolymerized PDMS has been shown not to compromise the functionality of the protein[46]. To test this observation with the newly developed inclusion method, solutions of laccase with concentrations ranging from 0 to 3.2 mg/ml were included as droplets in PDMS. Enzymatic activity was measured after two days and compared to an identical untreated control enzyme solution. To reliably perform a one-to-one comparison of residual activity between the control solution and the one stored as inclusions, we simply extracted the inclusion solution using a micropipette. The laccase solution included in PDMS retained its activity and was not influenced by the inclusion process, i.e. the residual activity measured from the included enzymatic solution was comparable to the untreated solutions (Figure 4d). The highest level of protein adsorption is usually registered at conditions of pH values close to the isoelectric point (pI) of the protein[47]. Since the laccase solution was prepared in 50% glycerol with the addition of 100 mM of potassium phosphate giving a pH of 7.5, a condition far from the pI of the protein, it is not surprising that the inclusion of laccase (pI = 3.535) within PDMS did not lead to a loss of enzymatic activity (Figure 4d). In addition, the high concentration of amphiphilic glycerol present in the droplet will also favour the retention of the biomolecules in the liquid phase, as glycerol can shield the interaction between the hydrophobic PDMS surface and the hydrophobic regions of the proteins.

The possibility of completely embedding liquid droplets in a material paves the way for the development of microreactors and sensing droplets, e.g. containing traceable elements whose behavior can be influenced by external triggers. By including solutions containing superparamagnetic nanoparticles, the localization and orientation of these particles could reveal the presence of externally applied magnetic fields, e.g. a magnet influenced their movement. Similarly, nanoparticle movement could be actively induced to swirl the solution within the droplet (Figure 4e) and clearly indicates that a liquid phase is maintained in the inclusions introduced in the PDMS matrix. These observations suggest applications of included nanoparticles as stirrers for microliter volumes and/or as drivers of the motion of droplets on surfaces[48–50].

**Conclusions**

We report a method for the complete inclusion of aqueous droplets containing dyes, biomolecules or nanoparticles in PDMS under mild conditions that do not alter their biochemical properties or mobility, respectively. This new tool allows to create liquid phases into a polymeric matrix, a process that will be useful for the development of functional materials from photonic crystals, batteries, biosensors, storage devices, microreactors to transdermal drug delivery systems which are under investigation in forthcoming works[51].

**Conflicts of interest**

Parts of this work were implemented in a European Patent Application (EP17197227.6;50617EP).

**References**

1. Song, H.; Tice, J. D.; Ismagilov, R. F. A Microfluidic System for Controlling Reaction Networks in Time. Angewandte Chemie 2003, 115, 792-796.

2. Bongsoo, K.; SangHyuk, Y.; Kim Yong-Jin; Jaeyoon, P.; Byunghoon, K.; Seungjoo, H.; Kang Sun-Woong; Keonwook, K.; Unyong, J. A Strain-Regulated, Refillable Elastic Patch for Controlled Release. Adv. Mater. Interfaces 2016, 3, 1500803.

3. Larmagnac, A.; Eggenberger, S.; Janossy, H.; Vörös, J. Stretchable electronics based on Ag-PDMS composites. Sci. Rep. 2014, 4, 7254.

4. Style, R. W.; Boltyanskiy, R.; Allen, B.; Jensen, K. E.; Foote, H. P.; Wettlaufer, J. S.; Dufresne, E. R. Stiffening solids with liquid inclusions. Nat Phys 2015, 11, 82-87.

5. Kim, H. K.; Lee, D.; Lim, S. Wideband-Switchable Metamaterial Absorber Using Injected Liquid Metal. Scientific Reports 2016, 6, 31823.

6. S. Khan; S. Tinku; L. Lorenzelli; R. S. Dahiya Flexible Tactile Sensors Using Screen-Printed P(VDF-TrFE) and MWCNT/PDMS Composites. IEEE Sensors Journal 2015, 15, 3146-3155.

7. Sadasivuni, K. K.; Mohiuddin, M.; Gao, X.; Akther, A.; Mun, S.; Kim, J., In Cellulose/PDMS hybrid material for actuating lens; 2015; Vol. 9434, pp 94340K-94340K-6.

8. Wang, W.; Rodrigue, H.; Ahn, S. Deployable Soft Composite Structures. Scientific Reports 2016, 6, 20869.

9. Martinez, R. V.; Branch, J. L.; Fish, C. R.; Lihua, J.; Shepherd, R. F.; Nunes, R. M.; Zhigang, S.; Whitesides, G. M. Robotic Tentacles with Three-Dimensional Mobility Based on Flexible Elastomers. Adv Mater 2012, 25, 205-212.

10. Iino, R.; Matsumoto, Y.; Nishino, K.; Yamaguchi, A.; Noji, H. Design of a large-scale femtoliter droplet array for single-cell analysis of drug-tolerant and drug-resistant bacteria. Frontiers in Microbiology 2013, 4.

11. Zhu, Y.; Zhu, L.; Guo, R.; Cui, H.; Ye, S.; Fang, Q. Nanoliter-Scale Protein Crystallization and Screening with a Microfluidic Droplet Robot. Scientific Reports 2014, 4, 5046.

12. Walt, D. R. Protein measurements in microwells. Lab Chip 2014, 14, 3195-3200.

13. Bose, S.; Wan, Z.; Carr, A.; Rizvi, A. H.; Vieira, G.; Pe'er, D. Scalable microfluidics for single cell RNA printing and sequencing. Genome Biol. 2015, 16, 120.

14. Klein, A. M.; Mazutis, L.; Akartuna, I.; Tallapragada, N.; Veres, A.; Li, V. Droplet barcoding for single-cell transcriptomics applied to embryonic stem cells. Cell 2015, 161, 1187-1201.

15. Lach, S.; Yoon, S. M.; Grzybowski, B. A. Tactic, reactive, and functional droplets outside of equilibrium. Chem. Soc. Rev. 2016, 45, 4766-4796.

16. Boukellal, H.; Selimovic, S.; Jia, Y.; Cristobal, G.; Fraden, S. Simple, robust storage of drops and fluids in a microfluidic device. Lab Chip 2009, 9, 331-338.

17. Stone, H. A.; Stroock, A. D.; Ajdari, A. ENGINEERING FLOWS IN SMALL DEVICES. Annu. Rev. Fluid Mech. 2004, 36, 381-411.

18. Song, H.; Chen, D. L.; Ismagilov, R. F. Reactions in Droplets in Microfluidic Channels. Angewandte Chemie International Edition 2006, 45, 7336-7356.

19. Loste E., Park R. J., Warren J., Meldrum F.C., Precipitation of Calcium Carbonate in Confinement. Advanced Functional Materials 14(12):1211-1220.

20. Kobayashi, T., Konishi, S.; Microfluidic chip with serially connected filters for improvement of collection efficiency in blood plasma separation. Sensors and Actuators B: Chemical 2012161(1), 1176-1183.

21. Jin, B.; Esteva-Font, C.; Verkman, A. S. Droplet-based microfluidic platform for measurement of rapid erythrocyte water transport. Lab Chip 2015, 15, 3380-3390.

22. Taylor, N.; Elbaum-Garfinkle, S.; Vaidya, N.; Zhang, H.; Stone, H. A.; Brangwynne, C. P. Biophysical characterization of organelle-based RNA/protein liquid phases using microfluidics. Soft Matter 2016, 12, 9142-9150.

23. Wang, J.; Liu, S.; Nahata, A. Reconfigurable plasmonic devices using liquid metals. Opt. Express 2012, 20, 12119-12126.

24. Liang, S.; Li, Y.; Chen, Y.; Yang, J.; Zhu, T.; Zhu, D.; He, C.; Liu, Y.; Handschuh-Wang, S.; Zhou, X. Liquid metal sponges for mechanically durable, all-soft, electrical conductors. J. Mater. Chem. C 2017, 5, 1586-1590.

25. Chipara, A. C.; Owuor, P. S.; Sanjit, B.; Gustavo, B.; Asif, S. A. S.; Mircea, C.; Robert, V.; Jun, L.; Galvao, D. S.; Tiwary, C. S.; Ajayan, P. M. Structural Reinforcement through Liquid Encapsulation. Adv. Mater. Interfaces 2017, 4, 1600781.

26. Lawton, R. A.; Price, C. R.; Runge, A. F.; Doherty III, W. J.; Saavedra, S. S. Air plasma treatment of submicron thick PDMS polymer films: effect of oxidation time and storage conditions. Colloids Surf. Physicochem. Eng. Aspects 2005, 253, 213-215.

27. Schneider, F.; Draheim, J.; Kamberger, R.; Wallrabe, U. Process and material properties of polydimethylsiloxane (PDMS) for Optical MEMS. Sensors and Actuators A: Physical 2009, 151, 95-99.

28. Mata, A.; Fleischman, A. J.; Roy, S. Characterization of Polydimethylsiloxane (PDMS) Properties for Biomedical Micro/Nanosystems. Biomed. Microdevices 2005, 7, 281-293.

29. Morteza, A.; Sahar, S.; Nelson, B. J.; Metin, S. Recent Advances in Wearable Transdermal Delivery Systems. Adv Mater 2018, 30, 1704530.

30. Wu, D.; Zhao, B.; Dai, Z.; Qin, J.; Lin, B. Grafting epoxy-modified hydrophilic polymers onto poly(dimethylsiloxane) microfluidic chip to resist nonspecific protein adsorption. Lab Chip 2006, 6, 942-947.

31. Roach, L. S.; Song, H.; Ismagilov, R. F. Controlling Nonspecific Protein Adsorption in a Plug-Based Microfluidic System by Controlling Interfacial Chemistry Using Fluorous-Phase Surfactants. Anal. Chem. 2005, 77, 785-796.

32. Faccio, G. From Protein Features to Sensing Surfaces. Sensors 2018, 18.

33. Liu, Y.; Zhang, L.; Wu, W.; Zhao, M.; Wang, W. Restraining non-specific adsorption of protein using Parylene C-caulked polydimethylsiloxane. Biomicrofluidics 2016, 10, 024126.

34. Chandradoss, S. D.; Haagsma, A. C.; Lee, Y. K.; Hwang, J.; Nam, J.; Joo, C. Surface Passivation for Single-molecule Protein Studies. Journal of Visualized Experiments: JoVE 2014, 50549.

35. Keir, B.; Wu Min-Hsien; Zheng, C.; Zhangfeng, C.; Watts, J. F.; Baker, M. A. Simple surface treatments to modify protein adsorption and cell attachment properties within a poly(dimethylsiloxane) micro-bioreactor. Surf. Interface Anal. 2006, 38, 198-201.

36. Herrmann, M.; Roy, E.; Veres, T.; Tabrizian, M. Microfluidic ELISA on non-passivated PDMS chip using magnetic bead transfer inside dual networks of channels. Lab Chip 2007, 7, 1546-1552.

37. McDonald, J. C.; Whitesides, G. M. Poly(dimethylsiloxane) as a Material for Fabricating Microfluidic Devices. Acc. Chem. Res. 2002, 35, 491-499.

38. Vagenende, V.; Yap, M. G. S.; Trout, B. L. Mechanisms of Protein Stabilization and Prevention of Protein Aggregation by Glycerol. Biochemistry (N. Y.) 2009, 48, 11084-11096.

39. Stepanenko, O. V.; Stepanenko, O. V.; Kuznetsova, I. M.; Verkhusha, V. V.; Turoverov, K. K. Beta-Barrel Scaffold of Fluorescent Proteins: Folding, Stability and Role in Chromophore Formation. International review of cell and molecular biology 2013, 302, 221-278.

40. Quiquampoix H., A stepwise approach to the understanding of extracellular enzyme activity in soil. I. Effect of electrostatic interactions on the conformation of a beta-D-glucosidase adsorbed on different mineral surfaces. Biochimie, 1987, Jun-Jul; 69(6-7):753-63

41. Baugh L., Vogel V., Structural changes of fibronectin adsorbed to model surfaces probed by fluorescence resonance energy transfer, J. Biomed. Mater. Res. A, 2004, Jun 1; 69(3):525-34

42. Chumbimuni-Torres, K.; Coronado, R. E.; Mfuh, A. M.; Castro-Guerrero, C.; Silva, M. F.; Negrete, G. R.; Bizios, R.; Garcia, C. D. Adsorption of proteins to thin-films of PDMS and its effect on the adhesion of human endothelial cells. RSC Adv. 2011, 1, 706-714.

43. Cai, Z.; Qiu, W.; Shao, G., Wang, W.; A new fabrication method for all-PDMS waveguides, Sensors and Actuators A: Physical 2013 204 44-47.

44. Rodríguez Couto, S.; Toca Herrera, J. L. Industrial and biotechnological applications of laccases: A review. Biotechnol. Adv. 2006, 24, 500-513.

45. Norde, W. My voyage of discovery to proteins in flatland …and beyond. Colloids and Surfaces B: Biointerfaces 2008, 61, 1-9.

46. Piscitelli, A.; Pezzella, C.; Giardina, P.; Faraco, V.; Sannia, G. Heterologous laccase production and its role in industrial applications. Bioengineered Bugs 2010, 1, 252-262.

47. Baldrian, P. Fungal laccases - occurrence and properties. FEMS Microbiol. Rev. 2006, 30, 215-242.

48. Heyries, K. A.; Marquette, C. A.; Blum, L. J. Straightforward Protein Immobilization on Sylgard 184 PDMS Microarray Surface. Langmuir 2007, 23, 4523-4527.

49. van Reenen, A.; de Jong, A. M.; den Toonder, Jaap M. J.; Prins, M. W. J. Integrated lab-on-chip biosensing systems based on magnetic particle actuation - a comprehensive review. Lab Chip 2014, 14, 1966-1986.

50. Chong, W. H.; Chin, L. K.; Tan, R. L. S.; Wang, H.; Liu, A. Q.; Chen, H. Stirring in Suspension: Nanometer-Sized Magnetic Stir Bars. Angewandte Chemie International Edition 2013, 52, 8570-8573.

51. Long, Z.; Shetty, A. M.; Solomon, M. J.; Larson, R. G. Fundamentals of magnet-actuated droplet manipulation on an open hydrophobic surface. Lab Chip 2009, 9, 1567-1575.

52. Innocenti Malini R., Lesage J., Toncelli C., Fortunato G., Rossi R. M., Spano F., Crosslinking dextran electrospun nanofibers via borate chemistry: Proof of concept for wound patches, European Polymer Journal, 110 (2019), 276-282.

# Supporting Information

**Complete inclusion of bioactive molecules and particles in polydimethylsiloxane: a straightforward process for the under mild conditions**

Greta Faccio[a*], Alice Cont[b], Erik Mailand[a], Elaheh Zare-Eelanjegh[a],

Riccardo Innocenti Malini[b], Katharina Maniura-Weber[a], René Michel Rossi[b], Fabrizio Spano[b*]

[a] Empa, Swiss Federal Laboratories for Materials Science and Technology, Laboratory for Biointerfaces, Lerchenfeldstrasse 5, CH-9014 St. Gallen, Switzerland.

[b] Empa, Swiss Federal Laboratories for Materials Science and Technology, Laboratory for Biomimetic Membranes and Textiles, Lerchenfeldstrasse 5, CH-9014 St. Gallen, Switzerland

*E-mail: fabrizio.spano@zhaw.ch

greta.faccio@gmail.com

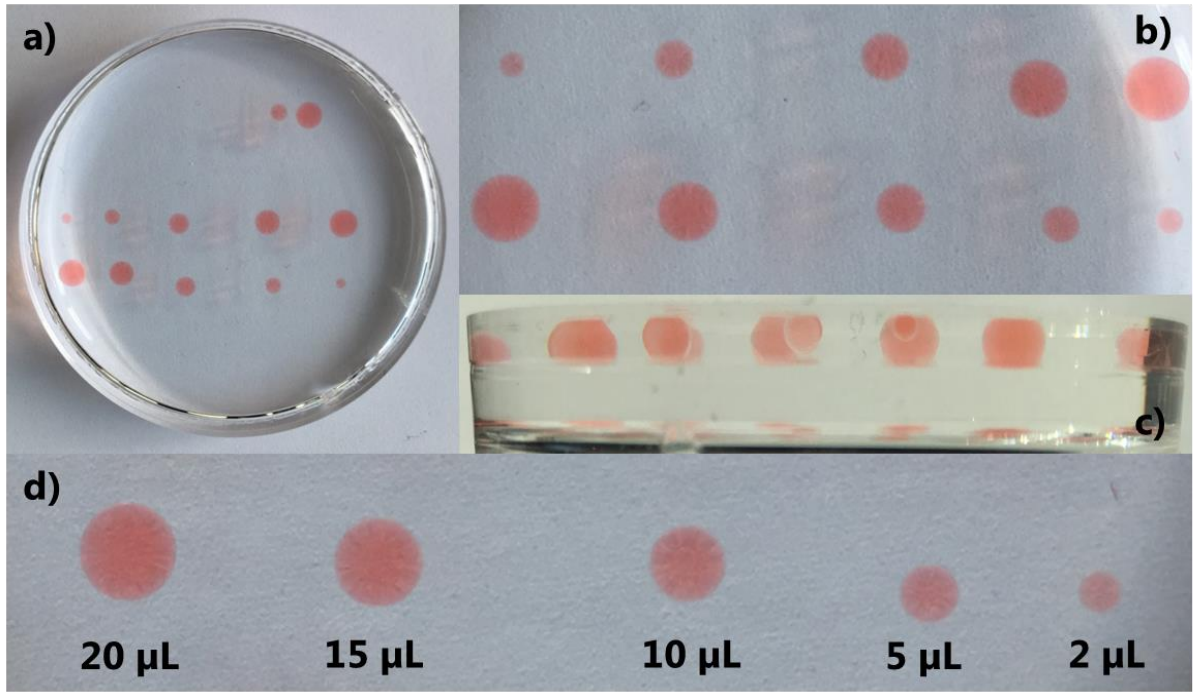

**Figure SI1.** Soft liquid composites sample made with inclusions of red dye loaded droplets: a) Top view of the sample; b) Zoomed view of the soft liquid composite sample illustrating the different droplet volumes; c) 90° side view of the sample illustrating the penetration of the red liquid droplets into the PDMS matrix. d) Zoomed top view illustrating the droplet volumes with 20, 15, 10, 5 and 2 µl, respectively.

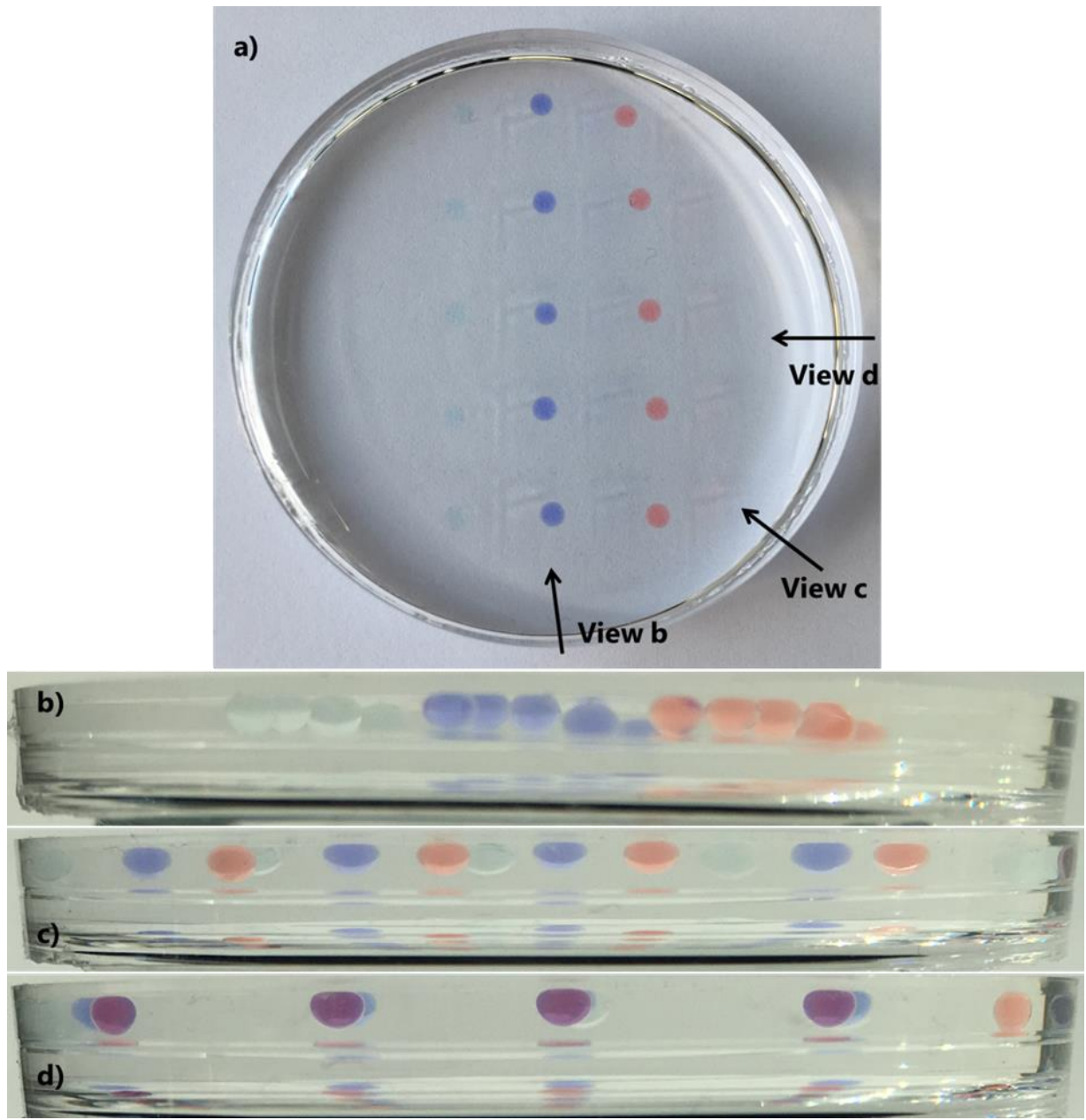

**Figure SI2.** Soft liquid composites sample made with inclusions using different dye-loaded droplets: a) Top view of the sample; b), c) and d) 90° side views of dye-loaded inclusions into the PDMS matrix changing colors due the superposition of the droplets in function of the selected lateral view angle.

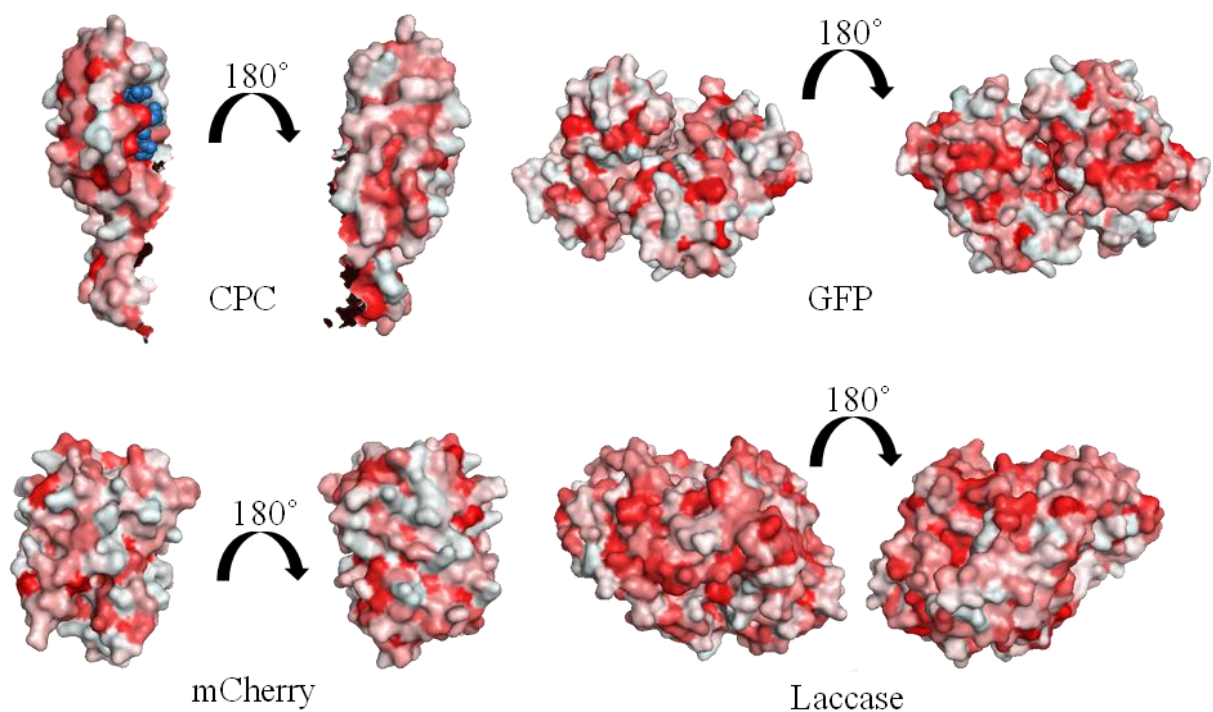

**Figure SI3.** Surface hydrophobicity of the proteins used in the study: CPC (with the bilin co-factor as blue spheres, PDB:4F0T), GFP (PDB:1EME), mCherry (PDB:2H5Q), and laccase (PDB:1KYA) visualized according to the normalized Eisenberg hydrophobicity scale showing hydrophilic regions as white and hydrophobic ones in red.

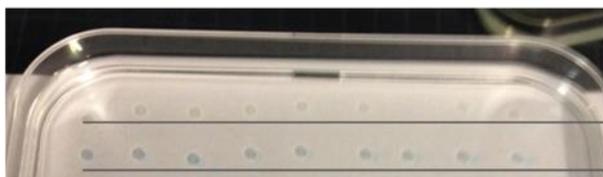

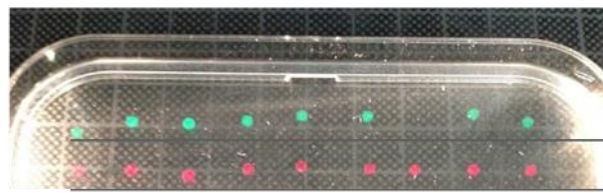

**Figure SI4.** Photographic imaging of fluorescent proteins GFP and CPC as liquid droplets in PDMS upon top and lateral illumination.